\documentclass[12pt]{article}
\newcommand{\bea}{\begin{eqnarray}}
\newcommand{\eea}{\end{eqnarray}}
\newcommand{\F}{\mathcal F}
\newcommand{\bH}{\mathbf H}
\newcommand{\id}{\mathbf{1}}
\newcommand{\DeltaF}{\Delta^{\mathcal F}}
\newcommand{\mH}{\mathcal H}
\newcommand{\llangle}{\langle \langle}
\newcommand{\rrangle}{\rangle \rangle}
\newcommand{\Gd}{\mathcal{G}_d}
\begin{document}
\renewcommand{\thefootnote}{\fnsymbol{footnote}}

\thispagestyle{empty}

\title{Snyder Noncommutativity and Pseudo-Hermitian Hamiltonians from a Jordanian Twist}
\author{P. G. Castro\thanks{{\em e-mail: pgcastro@cbpf.br}},~
R. Kullock\thanks{{\em e-mail: ricardokl@cbpf.br}}
 ~and F. Toppan\thanks{{\em e-mail: toppan@cbpf.br}}
\\ 
\\
{\it $~^{\ast}$ DM/ICE/UFJF, Campus Universit\'ario,} \\{\it cep 36036-330, Juiz de Fora (MG), Brazil.}
\\
{\it $~^{\dagger\ddagger}$ TEO/CBPF, Rua Dr.} {\it Xavier Sigaud 150,} \\{\it cep 22290-180, Rio de Janeiro (RJ), Brazil.}}
\maketitle
\begin{abstract}
Nonrelativistic quantum mechanics and conformal quantum mechanics are deformed through a Jordanian twist. The deformed space coordinates satisfy the Snyder noncommutativity.  The resulting deformed Hamiltonians are pseudo-Hermitian Hamiltonians of the type discussed by Mostafazadeh.\par
The quantization scheme makes use of the so-called ``unfolded formalism" discussed in previous works. A Hopf algebra structure, compatible with the physical interpretation of the coproduct, is introduced for the Universal Enveloping Algebra of a suitably chosen dynamical Lie algebra (the Hamiltonian is contained among its generators).\par
The multi-particle sector, uniquely determined by the deformed $2$-particle Hamiltonian, is composed of bosonic particles.
\end{abstract}
\vfill
\rightline{}
\rightline{CBPF-NF-006/11}
\newpage
\section{Introduction}
In this work we deform a non-relativistic quantum mechanical system through the Jordanian twist (a special case \cite{dv,ohn,og} of a non-abelian Drinfel'd twist \cite{drin85,drin88,res} applied to the Universal Enveloping of the $sl(2)$ algebra and such that its deformation parameter is dimensional).
We make use of what can be named the ``unfolded quantization framework", 
which has been at first discussed in \cite{ckt} and further elaborated in \cite{cckt}.  
We briefly recall its  basic tenets: at first an abstract dynamical Lie algebra ${\cal G}_d$ 
is introduced. It contains generators which are later identified with relevant physical 
operators (such as the Hamiltonian) and has the Heisenberg algebra as a subalgebra.  
The ${\cal U}({\cal G}_d)$ universal enveloping algebra possesses a Hopf-algebra 
structure with a correct physical interpretation of the undeformed coproduct, 
in particular the additivity of the energy when constructing undeformed 
multi-particle states (see also \cite{cct}). The universal enveloping Lie algebra is deformed via a 
Drinfel'd twist (the relevant formulas for twist-deformed Hopf algebra structures 
and costructures, namely generators, brackets, unit, counit, antipode,  multiplication, 
coproduct, etc.,  are reported in \cite{cckt} and references therein and will not be repeated here). 
In the case here considered, since the Jordanian twist involves the $sl(2)$ generators, 
$sl(2)$ has to be taken as a subalgebra of ${\cal G}_d$. 
\par
The single-particle deformed operators are recovered from the twist-deformed generators 
${g}^{\cal F}$ (${g}^{\cal F}\in {\cal U}({\cal G}_d)$ is a twist-deformation of the generator
$g\in {\cal G}_d$).  Ordinary brackets are taken among the deformed generators
(applying this scheme for an abelian-twist to the position operators $x_i$ one induces a Bopp-shift ${x_i}^{\cal F}$ and a constant noncommutativity $\Theta_{ij}$ obtained from $[{x_i}^{\cal F},{x_j}^{\cal F}]=\Theta_{ij}$ \cite{ckt, cckt, cct}).
\par
For what concerns the multi-particle operators, they will be constructed as coproducts of the deformed generators (more on that later).\par
Once we have identified, thanks to the Hopf-algebra structure, the relevant deformed single-particle (belonging to $ {\cal U}({\cal G}_d$)) and multi-particle generators (belonging
to ${\cal U}({\cal G}_d)\otimes\ldots \otimes {\cal U}({\cal G}_d)$), 
we can realize them as operators acting on the Hilbert space defined by the Fock space of the creation and annihilation operators. At this stage, the generators in ${\cal G}_d$ and their deformations  can be identified with composite operators 
constructed in terms of the creation and annihilation operators. This identification only holds
at the Lie algebra level and not within the Hopf algebra because it is spoiled by the coproduct
(for this reason one is forced to introduce and work with the Hopf algebra structure of the Universal Enveloping
Algebra of the ``unfolded" Lie algebra ${\cal G}_d$ and not of the Universal Enveloping Algebra of the Heisenberg algebra).\par
This general scheme was applied in \cite{cckt} for the abelian twist of the $d$-dimensional harmonic oscillator (producing a constant non-commutativity). We recall some of the relevant features one encounters.
The  deformed $2$-particle Hamiltonian, given by the undeformed coproduct of the deformed Hamiltonian (${{\bf H}^{\cal F}}_{12}\equiv \Delta ({\bf H}^{\cal F})$), is no longer additive
(${{\bf H}^{\cal F}}_{12}={{\bf H}^{\cal F}}_{1}+{{\bf H}^{\cal F}}_{2} +\Omega_{12}$). An extra term $\Omega_{12}$, which vanishes in the limit
for the deformation parameter going to zero, is induced  by the deformation.  The coassociativity
of the coproduct implies the associativity of the deformed multi-particle Hamiltonian (for the $3$-particle Hamiltonian ${{\bf H}^{\cal F}}_{123}$, the merging of the first particle with the combination of the second and of the third particle produces the same output as the merging of the third particle
with the combination of the first and the second particle, i.e.  ${{\bf H}^{\cal F}}_{123}= {{\bf H}^{\cal F}}_{(12)3} ={{\bf H}^{\cal F}}_{1(23)}$).\par

The deformed multi-particle Hamiltonians are symmetric in the exchange of particles (the particles behave as ordinary bosons even in presence of the deformation). 
The convenience of using the undeformed coproduct $\Delta ({\bf H}^{\cal F})$ of the deformed
Hamiltonian, instead of its deformed coproduct $\Delta^{\cal F} ({\bf H}^{\cal F})$, lies on the fact that in the latter case the symmetry under permutation is guaranteed under a twisted flip and not an ordinary flip \cite{bal,cckt}. The deformed coproduct is related to the undeformed coproduct via a twist
$\Delta^{\cal F} ({\bf H}^{\cal F})= \F\Delta ({\bf H}^{\cal F})\F^{-1}$. Applied to the Hilbert space,
due to the nature of the twist operator $\F$, deformed and undeformed coproducts are related by a unitary transformation (in the appendix we explicitly prove that deformed and undeformed coproducts are unitarily related also when applied to $n$-particle sectors for $n>2$).\par
The above properties hold true even for the Jordanian deformation. Furthermore,
highly non-trivial features are specific of the Jordanian twist. At first we should point out that the Jordanian twist can be applied to the harmonic oscillator and/or to the conformal quantum mechanical system, with or without a Calogero-type
potential, no matter whether the oscillatorial damping due to the DFF trick is present or not,  see
\cite{dff, papa, stro}.
In the presence of a Calogero potential the dynamical Lie algebra ${\cal G}_d$ becomes infinite-dimensional because it has to be complemented by newer and newer ``unfolded generators".\par
Under the Jordanian twist, the commutator between the deformed position operators gets the same non-commutativity as the one originally introduced by Snyder in \cite{snyder} (for the spatial sector).
An interesting and unexpected output is that we find a Jordanian-twist interpretation for the Snyder noncommutativity. Applications of this kind of noncommutativity have been studied, for instance, in \cite{bks,gp,leiva,mmss}.
\par
The Jordanian-deformed Hamiltonian ${\bf H}^{\cal F}$, unlike its abelian-deformed counterpart,
is no longer Hermitian. It falls however in a well-known class of Hamiltonians (the $\eta$-pseudo-Hermitian Hamiltonians) discussed by Mostafazadeh (see \cite{most} and references therein). The Hamiltonians of this class,
satisfying the pseudo-hermiticity condition
\bea
{{\bf H}^{\cal F}}^\dagger &=& \eta {\bf H}^{\cal F}\eta^{-1},
\eea
with $\eta$ a Hermitian operator, admit a real spectrum and can be mapped into Hermitian
operators by conveniently choosing the inner product of their Hilbert space.
The Jordanian deformation of the quantum mechanical system is a well-defined operation producing a consistent quantum-mechanical theory.
In the context of noncommutative theories, the arising of pseudo-Hermitian Hamiltonians of the above type has been observed in \cite{bf,fgs}.
The treatment of the pseudo-hermiticity in these works follows the same line as the one adopted here. Differently from our case, their noncommutativity is not derived from  a Jordanian twist.\par 
The scheme of the paper is the following. In Section {\bf 2} we introduce the dynamical Lie algebra ${\cal G}_d$ for the harmonic oscillator and/or the conformal quantum mechanics, which contains $sl(2)$ and the $d$-dimensional Heisenberg algebra as subalgebras. 
In Section {\bf 3} we extend the Jordanian deformation of ${\cal U}(sl(2))$ to a Jordanian deformation for ${\cal U}({\cal G}_d)$. We present the deformed generators (including the Hamiltonian) and the commutators among deformed position generators recovering, in particular, the Snyder noncommutativity.  In Section {\bf 4} we discuss the pseudo-hermiticity properties of the
deformed single-particle Hamiltonian. In Section {\bf 5} we compute the 
Jordanian-deformed Hamiltonians for the multi-particle sector (explicitly for $n=2, 3$).
In the Appendix we present the proof of the unitary equivalence between deformed and undeformed coproduct. In the Conclusions we present some outlooks of our results.

\section{On harmonic oscillators, conformal quantum mechanics and their dynamical Lie algebras.}

As in \cite{cckt}, we start with the Heisenberg algebra $h_d$ comprising generators $x_i$, $p_i$ and $\hbar$ (the central charge) satisfying
\begin{eqnarray}\label{heis}
[x_i,p_j]=i\hbar \delta_{ij}, &&  [\hbar,x_i]=[\hbar,p_i]=0,
\end{eqnarray}
with $i=1,\ldots,d.$ 

We enlarge this algebra by introducing the extra generators $H$, $K$ and $D$, in order to obtain a Lie algebra $\Gd$:
\begin{equation}
 {\cal{G}}_d = \{ \hbar, x_i, p_i,  H, K, D \}, 
\end{equation} 
which contains $h_d$ and $sl(2)$ as subalgebras.

The commutation relations of $\Gd$ are consistently recovered from the Heisenberg algebra after setting
\begin{eqnarray}{\label{ident}}
 H &=& \frac{1}{2\hbar} \left(p_ip_i \right) \nonumber\\ 
K &=& \frac{1}{2\hbar} \left(x_ix_i \right) \nonumber\\
 D&=& \frac{1}{4\hbar} \left(x_ip_i + p_ix_i \right).
\end{eqnarray}

As an abstract Lie algebra, $\Gd$ admits the following nonvanishing commutation relations:
\begin{eqnarray}{\label{Gd}}
\relax [x_i,p_j]&=&i\hbar \delta_{ij}\nonumber\\
\relax [D,H ] &= & iH\nonumber\\
\relax [D,K] &=& -iK \nonumber\\
\relax [K,H] &=& 2iD\nonumber \\
\relax [x_i,H] &=& ip_i\nonumber\\ 
\relax [x_i,D] &=& \frac{i}{2}x_i\nonumber \\ 
\relax [p_i,K] &=& -ix_i\nonumber\\ 
\relax [p_i,D] &=& -\frac{i}{2}p_i.
\end{eqnarray}

The Hamiltonian $\bH$ of the harmonic oscillator is a linear combination of $H$ and $K$, so that $\Gd$ can be regarded as the dynamical Lie algebra of the harmonic oscillator.

If we want to introduce, for instance, Calogero potentials, which are consistent with $sl(2)$-based quantum mechanics \cite{dff, papa, stro},  we need to further enlarge the algebra by introducing an extra generator $N$, whose commutation relations can be recovered from
\begin{equation}
N =\frac{1}{K} = \frac{2 \hbar }{x_ix_i}.
\end{equation}
This extra generator makes the algebra infinite-dimensional, since we now are obliged to introduce extra generators such as
\begin{equation}
N_i=\left[N,p_i\right],
\end{equation}
whose commutation relations are obtained by setting $N_i=-\frac{1}{K^2}[K,p_i]= - \frac{i  x_i}{K^2}$. This requires the introduction of a new generator
\begin{equation}
N_{ij}=\left[N_i, p_j\right],
\end{equation}
with commutation relations recovered from $N_{ij}= \frac{ \hbar \delta_{ij}}{K^2} - \frac{2  x_ix_j }{K^3}$,  and so on.

There exists an $sl(2)$ subalgebra within this larger, infinite-dimensional version of $\Gd$, expressed by $K$, $D$ and $H_g=H+gN$, with $g$ a coupling constant, satisfying
\begin{eqnarray}
\relax [D,H_g ] &= & iH_g\nonumber\\
\relax [D,K] &=& -iK \nonumber\\
\relax [K,H_g] &=& 2iD.
\end{eqnarray}

\section{The Jordanian twist of ${\cal U}(sl(2))$ and the Snyder noncommutativity.}

There are only two inequivalent deformations of $sl(2)$. The first one is the standard deformation depending on a non-dimensional parameter $q$ leading to the quantum group ${\cal U}_q(sl(2))$ \cite{kr, jimbo}, which cannot be obtained from the Drinfel'd twist technique. The second one, which we shall use here, is called the Jordanian deformation of $sl(2)$ \cite{dv, ohn, og, kul, tol}, and can be obtained from the twist
\begin{equation}
\F=\exp\left(-iD\otimes\sigma\right),
\end{equation}
where  $\sigma=\ln(\mathbf{1}+\xi H)$. The parameter $\xi$ is dimensional and is taken as a real, positive number. 

We shall apply this twist to the Universal Enveloping Algebra $\mathcal{U}(\Gd)$ described in the previous section. As usual (see \cite{cckt}), the twist induces a deformation $g\mapsto g^\F$ on the generators of $\Gd$, with $g^\F\in\mathcal{U}(\Gd)$. Explicitly, the deformed generators are
\bea
x_i^\F&=&x_i e^{\frac{\sigma}{2}}\nonumber\\
p_i^\F&=&p_i e^{-\frac{\sigma}{2}}\nonumber\\
H^\F&=&H e^{-\sigma}\nonumber\\
K^\F&=&K e^{\sigma},
\eea
the others remaining undeformed. 

As we have seen,  $\Gd$ is the dynamical Lie algebra for the harmonic oscillator, whose Hamiltonian is given by $\mathbf{H}=H+K$. We can therefore write its deformation as
\begin{equation}
\mathbf{H}^\F=H^\F+K^\F=H e^{-\sigma}+K e^{\sigma}.
\end{equation}

We now present the ordinary commutators of the deformed generators (this is called the ``hybrid formalism'', see \cite{ckt}). The commutator of the deformed position variables yields precisely the Snyder noncommutativity \cite{snyder}:
\bea
[x_i^\F, x_j^\F]&=&-\frac{i\xi}{2}(x_i^\F p_j^\F-x_j^\F p_i^\F).
\eea

The other nonvanishing commutators are
\bea
\relax[x_i^\F, p_j^\F]&=&i\hbar\delta_{ij}+\frac{i\xi}{2}p_i^\F p_j^\F \nonumber\\
\relax[x_i^\F, D^\F]&=&\frac{i}{2}(x_i^\F -\xi x_i^\F H^\F ) \nonumber\\
\relax[x_i^\F, H^\F]&=&ip_i^\F(\mathbf{1}-\xi H^\F) \nonumber\\
\relax[x_i^\F, K^\F]&=&-\frac{\xi}{2}x_i^\F\left(\mathbf{1}+\frac{\xi}{2} H^\F\right)+i\xi(K^\F p_i^\F+D^\F x_i^\F)  \nonumber\\
\relax[p_i^\F, D^\F]&=&-ip_i^\F\left(\mathbf{1}-\frac{\xi}{2} H^\F\right) \nonumber\\
\relax[p_i^\F, K^\F]&=&-i(x_i^\F+\xi p_i^\F D^\F)+\frac{\xi^2}{4}p_i^\F H^\F \nonumber\\
\relax[D^\F, H^\F]&=&iH^\F(\mathbf{1}-\xi H^\F) \nonumber\\
\relax[D^\F, K^\F]&=&-iK^\F(\mathbf{1}-\xi H^\F) \nonumber\\
\relax[K^\F,H^\F]&=&2iD^\F(\mathbf{1}+\xi H^\F)+2\xi H^\F-2\xi^2 (H^\F)^2.
\eea

Regarding the coproduct, as discussed in \cite{cckt} and shown in appendix \textbf{A}, the operators coming from the undeformed and deformed coproducts are unitarily equivalent for all $n$-particle states. We are therefore entitled to use the undeformed coproduct $\Delta(\bH^\F)$, which has the advantage of exhibiting manifest symmetry under particle exchange.

\section{$\eta$-Pseudo-Hermitian Hamiltonians}
Suppose we have a Hilbert space $\mH$ with an operator which is to be considered the Hamiltonian of the system. It is not always that we encounter a Hermitian operator for that role. Here, the deformed Hamiltonian has precisely this difficulty. Indeed,
\begin{equation}
\bH^{\F\dagger}=\eta\bH^\F\eta^{-1},
\end{equation}
where
\begin{equation}
\eta = e^\sigma=\mathbf{1}+\xi H.
\end{equation}

The treatment for the type of Hamiltonian found here is to define \cite{kato, most} an inner product $\llangle,\rrangle$ under which $\mathbf{H}^\F$ is self-adjoint. Obviously this deformed inner product needs to satisfy some conditions, for instance, positiveness.  

This different inner product is related to the usual one by
\begin{equation}\label{inner product}
\llangle \psi,\phi \rrangle=\langle \psi, \eta \phi \rangle.
\end{equation}
Note that $\eta$ is a Hermitian, linear and invertible operator, and is analogous to a metric in a usual finite-dimensional vector space. 

Although as a vector space the Hilbert space endowed with the $\eta$-deformed inner product is isomorphic to the original one, as Hilbert spaces this is not true. Thus, we denote the Hilbert space with the $\eta$-deformed inner product as $\tilde \mH$. If $\eta$ is positive definite, the new inner product will also be so. 
 
Under this inner product we have  
\begin{eqnarray}
\nonumber \llangle \psi, \mathbf{H}^\F \phi \rrangle &=& \langle \psi ,e^{\sigma} \mathbf{H}^\F \phi \rangle \\ 
\nonumber &=& \langle e^{\sigma} \mathbf{H}^\F e^{-\sigma} e^{\sigma} \psi, \phi \rangle \\
\nonumber &=&\langle \mathbf{H}^\F\psi, e^{\sigma}\phi \rangle \\
&=& \llangle \mathbf{H}^\F \psi, \phi \rrangle.
\end{eqnarray}
Therefore, under the deformed inner product the Hamiltonian becomes self-adjoint. Operators such as $\bH^\F$, which are self-adjoint under the $\eta$-deformed inner product, are called \emph{$\eta$-pseudo-Hermitian} operators \cite{most}. 

It remains to be seen if the spectrum of the $\eta$-pseudo-Hermitian Hamiltonian is real. This will be true because the formal square root of $\eta$, $\rho$ such that $\rho^2=\eta$, is a unitary transformation $\rho : \tilde \mH \rightarrow \mH$, even though it is a Hermitian operator in $\mH$ \cite{froi}. 

To see this, consider a linear transformation $T: W \rightarrow V$ between two finite-di\-men\-sional vector spaces $W$ and $V$. Its adjoint will then be a transformation $T^{\ddagger}: V^{\ast} \rightarrow W^{\ast}$, where $V^{\ast}$ and $W^{\ast}$ are duals to the original vector spaces. Then we have, by definition,
\begin{equation}
\langle w,T^{-1}v \rangle_W=\langle T^{-1 \ddagger}w,v \rangle_V,
\end{equation}
where $\langle,\rangle_W$ is the inner product in $W$ and $\langle,\rangle_V$ is the one in $V$. 

In our notation for the inner products of $\mH$ and $\tilde{\mH}$ we have, for $\rho=\exp{{\frac12 \sigma}}$,
\begin{equation}
\llangle \tilde \psi ,\rho^{-1} \phi \rrangle =\langle \rho^{-1 \ddagger}\tilde \psi, \phi \rangle,
\end{equation}
with $\tilde \psi \in \tilde \mH$ and $\phi \in \mH$. Using expression (\ref{inner product}) we  also find that
\begin{equation}
\llangle \tilde \psi ,\rho^{-1} \phi \rrangle =\langle \rho \tilde \psi, \phi \rangle.
\end{equation}
Therefore $\rho^{-1 \ddagger} = \rho$, so that $\rho$ is unitary when regarded as a transformation $\rho : \tilde \mH \rightarrow \mH$. Note that $\tilde \psi$ can also be regarded as a vector in $\mH$, since $\mH$ and $\tilde{\mH}$ are identified as vector spaces.

A useful way of dealing with this scenario is to map all observables on $\tilde \mH$ back onto $\mH$ where the inner product is the usual one. This is done by 
\begin{equation}
\mathbf{H}^\F \mapsto \mathbf{H}^\F_{\rho} = \rho \mathbf{H}^\F \rho^{-1}.
\end{equation}
The new Hamiltonian will be given by 
\begin{equation}
\mathbf{H}^{\F}_{\rho} =  \left( 1 - \frac{\xi^2}{4} \right)H^{\F}+ K^{\F} +i \xi D,
\end{equation}
which is explicitly Hermitian since $K^{\F \dagger} = K^{\F} + 2i\xi D$. For our present consideration, $\xi$ is a small parameter. This shows that our pseudo-Hermitian Hamiltonian is related to a manifestly Hermitian Hamiltonian by a unitary transformation, and is therefore guaranteed to have a real spectrum. The transformation $\rho$ is called a pseudo-canonical transformation; the systems described by $\bH^\F$ in $\tilde\mH$ and $\bH^\F_\rho$ in $\mH$ are clearly physically equivalent \cite{mostjmp}.

The Hamiltonian $\bH^\F_\rho$, expressed in terms of the position and momentum operators
via the identification (\ref{ident}) and after setting $\hbar=1$, reads as:
\begin{equation}
\mathbf{H}^{\F}_{\rho} =  \left( 1 - \frac{\xi^2}{4} \right)\frac{p_i p_i}{2+\xi p_ip_i}+ \frac{x_i x_i}{2}\left(1+\xi \frac{p_ip_i}{2}\right) +i \xi \frac{x_ip_i+p_ix_i}{4}.
\end{equation}

\section{The deformed multi-particle Hamiltonians}  

As anticipated in section \textbf{3}, we shall make use of the undeformed coproduct of the deformed Hamiltonian. We obtain
\bea\label{2part}
\Delta(\bH^\F)&=&Ke^\sigma\otimes e^\sigma+e^\sigma\otimes Ke^\sigma-\xi^2 (KH\otimes H+ H\otimes KH)   \nonumber\\
&&+\sum_{n=1}^\infty(-\xi)^{n-1}\sum_{k=0}^n {n \choose k}  H^k\otimes H^{n-k}
\eea
This can be written as ${{\bf H}^{\cal F}}_{12}={{\bf H}^{\cal F}}_{1}+{{\bf H}^{\cal F}}_{2} +\Omega_{12}$. The term $\Omega_{12}$ can be recovered from (\ref{2part}). We see from these expressions that the energy is no longer additive. The two-particle Hamiltonian can be reexpressed in terms of the position and momentum operators via the identification
(\ref{ident}) and after setting $\hbar =1$.

Regarding the three-particle states, the energy will still be associative. This comes from the coassociativity of the coproduct, 
\begin{equation}
\Delta_{(2)}(\mathbf{H}^\mathcal{F})=(id\otimes\Delta)\Delta(\mathbf{H}^\mathcal{F})=(\Delta\otimes id)\Delta(\mathbf{H}^\mathcal{F}),
\end{equation}
where, explicitly,
\bea
\Delta_{(2)}(\mathbf{H}^\mathcal{F})&=&[K\otimes\id\otimes\id+\id\otimes K\otimes\id+\id\otimes\id\otimes K][e^\sigma\otimes e^\sigma\otimes e^\sigma \nonumber\\ 
&&-\xi^2(H\otimes H\otimes\id+H\otimes\id\otimes H+\id\otimes H\otimes H)-\xi^4(H\otimes H\otimes H)]\nonumber\\
&&+\sum_{n=1}^\infty(-\xi)^{n-1}\sum_{k=0}^n {n \choose k}\sum_{l=0}^k {k \choose l} H^l\otimes H^{k-l}\otimes H^{n-k}
\eea

The symmetry under particle exchange is guaranteed. Even in the presence of the Jordanian deformation, the particles behave as usual bosons.

\section{Conclusions}
In this paper we introduced a Jordanian deformation for the $d$-dimensional harmonic oscillator and/or the conformal quantum mechanics. We used the so-called unfolded quantization formalism based on a Hopf algebra construction which guarantees the physical interpretation
of the coproduct. The Jordanian deformation is applied to a dynamical Lie algebra which contains,
as a subalgebra, both the $sl(2)$ and the Heisenberg algebras. 
\par
We proved that the Jordanian quantization induces a consistent deformation. The theory
consists of bosonic particles even in the presence of the deformation. The deformed multi-particle Hamiltonian is associative and determined by the non-additive term entering the deformed $2$-particle Hamiltonian. \par
An interesting and unexpected feature of the Jordanian deformation is the fact that it induces a noncommutativity of Snyder type for the deformed space coordinates. 
The deformed single-particle Hamiltonian is no longer Hermitian, but it belongs to the class of
so-called $\eta$-type pseudo-Hermitian Hamiltonians discussed by Mostafazadeh. Its spectrum
is real. The pseudo-Hermitian Hamiltonian is self-adjoint under an $\eta$-modified inner product, and can be mapped into a Hermitian Hamiltonian possessing the same spectrum. \par
The importance of the Jordanian deformation is that it can be applied to any system containing an $sl(2)$ subalgebra.
\appendix
\section{Appendix: on the unitary equivalence of operators based on the deformed coproduct and the undeformed coproduct}

It has already been proven that the twist establishes a unitary equivalence between the undeformed coproduct and its deformed counterpart. It is still left to be shown, however, that this is true also for higher tensor products, lifted by the use of the coproduct itself.

Assume the twist $\F=U_{1}=U$ to be unitary under the conjugation $(A \otimes B)^{\dagger}= A^{\dagger} \otimes B^{\dagger}$. The general expression for the deformed $(n+1)$-particle expression of an operator, written in terms of its $n$-particle one, is

\begin{equation}
\DeltaF_{(n+1)} = \left( \DeltaF \otimes id^{\otimes n} \right) \DeltaF_{(n)},
\end{equation}
where it has been shown \cite{sweedler} that it is equivalent to any other ordering of $\DeltaF$ and the $id$ identity maps. 

We now assume $\DeltaF_{(n)}=U_n\Delta_{(n)} U_n^{-1}$ for some unitary operator $U_n$. Then, the equation above becomes
\begin{equation}
\DeltaF_{(n+1)}  =  \F_{12} \left[ \left( \Delta \otimes id^{\otimes n} \right) \DeltaF_{(n)} \right] \F_{12}^{-1}.
\end{equation}
with $\F_{12}= f^{\alpha}\otimes f_{\alpha}\otimes \id \otimes ... \otimes \id$. This becomes

\begin{equation}
 \DeltaF_{(n+1)} = 
 \F_{12} \left[ \left( (\Delta \otimes id^{\otimes n} ) U_n \right) \left( (\Delta \otimes id^{\otimes n}) \DeltaF_{(n)}\right) \left( (\Delta \otimes id^{\otimes n}) U_n^{-1}\right) \right] \F_{12}^{-1}
\end{equation}

If $U_n = \sum_k \frac{i^k}{k!} (u_1 \otimes ... \otimes u_n)^k$, with each $u_i$ Hermitian, then 
\begin{equation}
\left( (\Delta \otimes id^{\otimes n} ) U_n \right)^{\dagger} =  \sum \frac{(-i)^n}{n!} (\Delta(u_1^k))^{\dagger} \otimes ... \otimes u_n^{k \dagger}.
\end{equation}

The proof will be completed if $ \Delta(u) $ is Hermitian when $u$ is. To see this, let us assume $u = v_1v_2...v_k $ for some finite $k$ (i.e., $u$ is not primitive, but still a monomial), where each $v_i$ is primitive. Then 
\begin{equation}
(\Delta(u))^{\dagger}= (\Delta(v_k))^{\dagger}...(\Delta(v_1))^{\dagger} = (\Delta(v_k^{\dagger}))...(\Delta(v_1^{\dagger})) = \Delta(v_k^{\dagger}...v_1^{\dagger}) = \Delta(u),
\end{equation}
since $(\Delta(v_i))^{\dagger} = v_i^{\dagger} \otimes \id + \id \otimes v_i^{\dagger}= \Delta(v_k^{\dagger})$ and $u^{\dagger} = v_k^{\dagger}...v_1^{\dagger}$. For $u$ not a monomial this proof generalizes trivially due to the linearity of the coproduct.

Thus, if $U_n$ is unitary with a formal expansion $U_n = \sum_k \frac{i^k}{k!} (u_1 \otimes ... \otimes u_n)^k$, then $U_{n+1}= \F_{12} ((\Delta \otimes id^{\otimes n}) U_n)$ will still be unitary. Since $U_1=\F$ is unitary this means all $n$-particle operators will be unitary.

Furthermore, in the present case we have $\F = \exp(-i D \otimes \sigma)$, with $D$ a primitive element. It is therefore easy to check that we have the explicit expression
\begin{equation}
U_n = \prod_{j=2}^{n+1} \left( \prod_{i=1}^{j-1} \F_{ij} \right).
\end{equation}

{}~
\\{}~
\par {\large{\bf Acknowledgments}}{} ~\\{}~\par
 P.\ G.\ C.\ acknowledges financial support from FAPEMIG. This work was supported by 
CNPq (R.\ K.\ and F.\ T.).  We acknowledge precious discussions with B. Chakraborty, Z. Kuznetsova and J. Zubelli.

\end{document}